\documentclass[runningheads]{llncs}
\usepackage{graphicx}
\usepackage{hyperref}
\usepackage{ifthen}
\usepackage{textcomp}
\usepackage{xcolor}
\usepackage{svg}
\usepackage{cite}
\usepackage{amsmath,amssymb,amsfonts}
\usepackage{algorithmic}
\usepackage{fancyvrb}
\usepackage{listings}

\usepackage[T1]{fontenc}
\newboolean{showcomments}
\setboolean{showcomments}{true} 
\ifthenelse{\boolean{showcomments}}
  {\newcommand{\nb}[2]{
    \fcolorbox{gray}{yellow}{\bfseries\sffamily\scriptsize#1}
    {\sf\small$\blacktriangleright$\textit{#2}$\blacktriangleleft$}
   }
   
  }
  {\newcommand{\nb}[2]{}
   
  }

\definecolor{codered}{RGB}{177,84,105}
\definecolor{codeblue}{RGB}{86,53,255}
\definecolor{codegreen}{RGB}{63,127,95}
\definecolor{codesalmon}{RGB}{250,128,114}

\begin{document}
\title{Vibe-driven model-based engineering}
\titlerunning{Vibe-driven Model-based Engineering}
%

\author{Jordi Cabot\inst{1,2}\orcidID{0000-0003-2418-2489}}

\authorrunning{J. Cabot}

%
\institute{Luxembourg Institute of Science and Technology, Esch-sur-Alzette, Luxembourg \\
\email{jordi.cabot@list.lu} \and
University of Luxembourg, Esch-sur-Alzette, Luxembourg }
\maketitle              
\begin{abstract}

There is a pressing need for better development methods and tools to keep up with the growing demand and increasing complexity of new software systems. 
New types of user interfaces, the need for intelligent components,  sustainability concerns, etc. bring new challenges that we need to handle. 
In the last years, model-driven engineering (MDE), including its latest incarnation, i.e. low/no-code development, has been key to improving the quality and productivity of software development, but models themselves are becoming increasingly complex to specify and manage. 
At the same time, we are witnessing the growing popularity of vibe coding approaches that rely on Large Language Models (LLMs) to transform natural language descriptions into running code at the expense of potential code vulnerabilities, scalability issues and maintainability concerns.

While many may think vibe coding will replace model-based engineering, in this paper we argue that, in fact, the two approaches can complement each other and provide altogether different development paths for different types of software systems, development scenarios, and user profiles.
In this sense, we introduce the concept of \textit{vibe-driven model-based engineering} as a novel approach to integrate the best of both worlds (AI and MDE) to accelerate the development of reliable  complex systems. We outline the key concepts of this new approach and highlight the opportunities and open challenges it presents for the future of software development.

\end{abstract}

\section{Introduction}

Current software development projects face a growing demand for advanced features, including support for new types of user interfaces (augmented reality, virtual reality, chat and voice,...), intelligent behavior to be able to classify/predict/recommend information based on user input or the need to face new security and sustainability concerns, among many other new types of requirements.   

To tame this complexity, software engineers typically used to choose to work at a higher abstraction level~\cite{Booch18} where technical details can be ignored, at least during the initial development phases.
{\em Low-code platforms} are the latest incarnation of this trend, promising to accelerate software delivery by dramatically reducing the amount of hand-coding required. 
Low-code can be regarded as a continuation or specific style of other model-based approaches~\cite{cabot2024lowcode, RuscioKLPTW22}, where high-level software models are used to (semi)automatically generate the running software system.
However, even models themselves are becoming more and more complex due to the increasing complexity of the underlying systems being modeled. 
Beyond ``classical'' data and behavioral aspects, we now need to come up with new models to define the new types of UIs or all the smart features of the system. 
This hampers the adoption of model-driven processes as it reduces the (perceived?) Return on Investment (ROI) of modeling activities due to the increased cost of modeling\footnote{Note that the adoption of modeling practices is a complex sociotechnical problem \cite{HutchinsonWR14}.}. 

In parallel, we are witnessing the explosion of vibe-coding, promising to generate full software applications from natural language descriptions thanks to the use of Large Language Models (LLMs)
\footnote{\url{https://en.wikipedia.org/wiki/Vibe_coding}}. 
With the new agentic capabilities provided by many IDEs and agent-based systems built on top of those LLMs, even the testing and verification of the generated code is becoming easier, where the agent itself creates tests, runs them and automatically refines the code if it detects any issues. And while the generated code is not always correct, it does keep improving with every new released LLM version.

So good (at first sight), that many developers are blindly adopting vibe-coding and disregarding more traditional (and reliable) approaches based on the use of software models and code-generation templates.
We even often see claims about the death of low-code with some of the major low-code commercial vendors reshaping their marketing strategies to present themselves as agentic platforms, hiding the fact that they still rely on software models and code-generation templates under the hood.

In this paper, we propose a new flexible development approach that explicitly combines the two strategies. We call it \textit{vibe-driven model-based engineering}. We say it is flexible because, based on the needs of the scenario and the expertise
of the user, we can choose to use the more traditional model-based approach relying on code-generation templates, the vibe-coding approach with LLMs for generating code or a combination of both. But the key aspect of our proposal is that regardless of which path you go, 
software models (or specifications or designs, depending on which terminology you prefer) remain the central point of the development process and guarantee a certain level of quality and reliability in the final software system.

The next sections are organized as follows. Section~\ref{sec:state-of-the-art} reviews the state of the art in applying LLMs and AI to development activities. 
Section~\ref{sec:vibe-driven-engineering} reviews the key concepts of our vibe-driven model-based engineering proposal. 
Then, Section~\ref{sec:variants} discusses the different variants of our approach depending on the needs of the scenario and the expertise of the user while Section \ref{sec:tool} comments on the infrastructure required to support vibe modeling. 
Finally, we discuss open challenges and future directions before concluding the paper.

\section{State of the art}
\label{sec:state-of-the-art}

The software engineering community has deeply embraced vibe coding and other LLM-based approaches to generate/review/test/evolve code from natural language descriptions. 
In fact, any major software engineering conference is now flooded with papers on this topic 
\footnote{You can check it yourself by looking at the proceedings of those conferences or any other bibliographic database. As anecdotal evidence, 70\% of the papers in arxiv are now LLM related, according to this post
\url{https://shape-of-code.com/2026/03/22/70-of-new-software-engineering-papers-on-arxiv-are-llm-related/}.}. 
At the same time, the community is also deeply reflecting on the implications of these new approaches and the role of LLMs (and AI in general) in software engineering. As one of the many examples of such reflecitons, see~\cite{TerragniVRB25} for a comprehensive review of the state of the art in AI-driven software engineering.

The modeling community has also embraced with real interest the idea of using LLMs to assist in modeling activities \cite{BurguenoCWZ22, Lola24, RoccoRSNR25} 
as part of ongoing efforts to reduce the cost of modeling itself and improve its ROI \cite{lowmodeling}. This is what we call \textit{vibe modeling}~\cite{vibemodeling}.
Many of these works focus on inferring a (partial) model from a natural language description. 
For instance, research conducted in~\cite{CamaraTBV23},~\cite{fill_conceptual_2023}, and~\cite{chaaben_towards_2023} evaluated the potential to create domain models from textual descriptions using prompting.  
Other approaches tried to go beyond simple prompting techniques and experiment with chain-of-thought \cite{chen2023automateddomainmodeling} and tree-of-thought \cite{Silva24} prompting techniques for better accuracy 
while still acknowledging the need for a human in the loop to validate the model and provide feedback to the LLM \cite{SilvaMCKP25}.

However, and while LLMs can clearly help in the modeling process, there is also the risk that developers believe that vibe coding can replace the need for models and code-generation templates altogether. 
As such the modeling community is also starting to reflect on their role in this AI-driven era. See, for instance, this manifesto \cite{manifestmodeling} that states that 
"to build a better world with AI, we must fundamentally rethink the partnership between human modelers and AI.".

Next section presents our concrete proposal to enable this partnership between modeling (including, but not only, human modelers) and AI a fruitful reality.

\section{Overview of vibe-driven model-based engineering}
\label{sec:vibe-driven-engineering}

\begin{figure*}
  \centering
  \includegraphics[width=\textwidth]{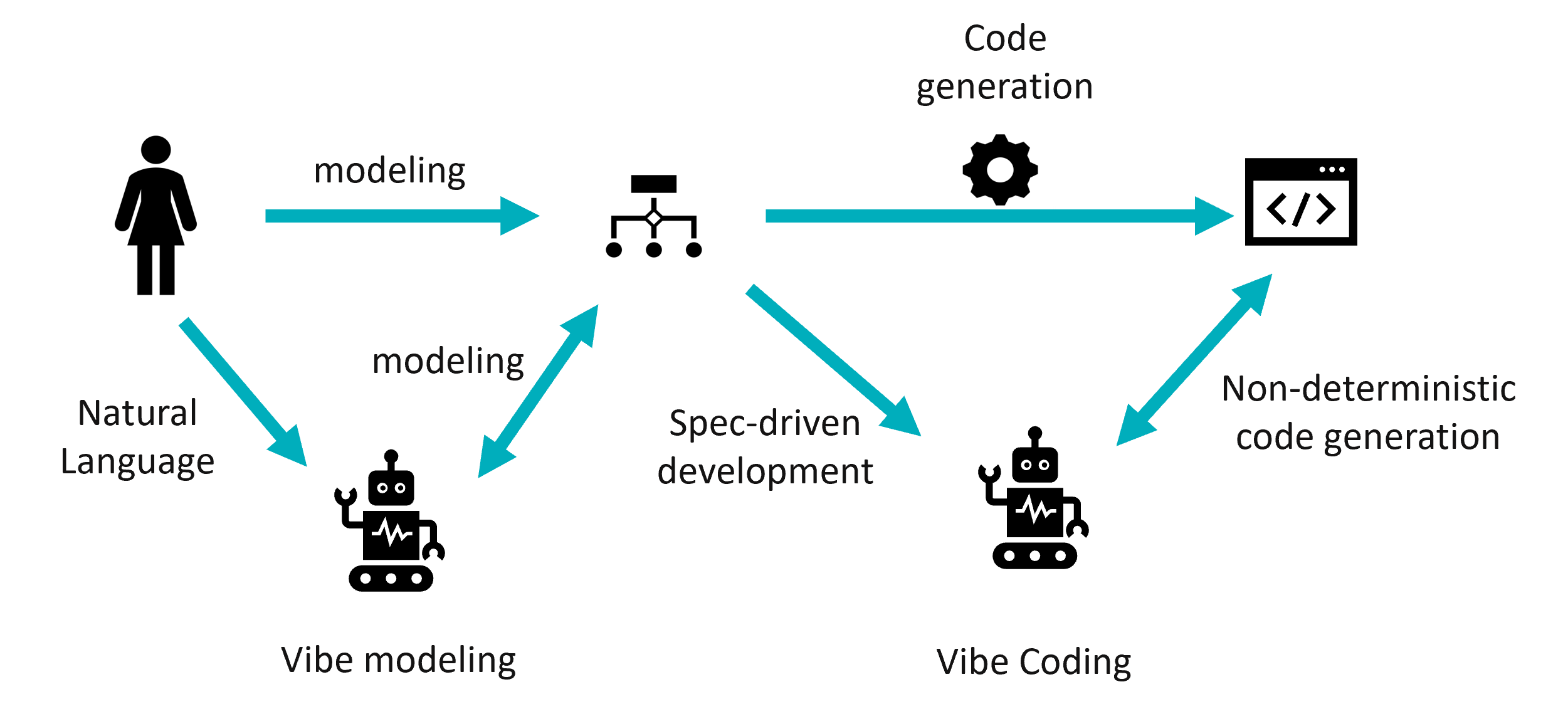}
  \caption{Possible vibe-driven model-based engineering development workflows.}
  \label{fig:vibe-driven-mbe-overview}
\end{figure*}

Figure~\ref{fig:vibe-driven-mbe-overview} summarizes how we see classical model-based development and LLM-assisted ``vibe'' workflows coexisting in a single, model-centric landscape.

The upper path recalls the classical model-driven/low-code process: a human models the system-to-be and picks the right rule-based code-generator for the target implementation platform. As a result, we obtain a completely deterministic software system. This system is completely deterministic and the code can be of high quality as the generators could embed “best practices” in the template, 
resulting in code that is secure, unbiased, energy efficient, etc and, more importantly, that doesn't need to be tested. If the models were correct, the code will be correct as well \footnote{Of course assuming that the code-generation templates work well but the templates can be exhaustively tested and even certified as they are reused across all projects targeting the same tech stack}.

This is in contrast with pure vibe coding approaches, where systems are generated from natural language descriptions and the result is unpredictable. Still, this does not mean we cannot benefit from the power of AI, e.g. to facilitate the modeling activity itself. It just means that we need to leverage AI in a controlled way.

Indeed, depending on the complexity of the system to be developed and their own skills and preferences, users can decide to opt for introducing vibe modeling in the development process. We can embed agents in the low-code platforms to assist the modelers. This speeds up the modeling process while still giving the users the chance to verify and validate the models before generating the code with the same rule-based deterministic generator as before. Note how this path combines the flexibility of AI and the determinism of rule-based code generation, offering a way to get the best of both worlds.

Still, introducing AI as a modeling assistant may not always be enough. Some scenarios may require a more flexible approach where even the code itself is generated by AI to cover unforeseen situations (a target platform not covered by the generator,
the need to generate code that goes beyond of the generator scope, addition of features that are not easy to model, e.g. styling visual aspects, etc). This is where vibe coding comes into play. 
But with a twist. In our approach, vibe coding does not start from a natural language description, but from a model. The model is used to guide the generation process. 

 It is worth noting that in every scenario, we keep the models as the pillar of the development. These models can be manually created or “vibed” but they are still explicit and can be reviewed and validated before generating (or vibing) the code. 
 Moreover, in the latter case, the models are part of the vibe coding input (a kind of “spec-driven development”, the term used by the vibe-coding community \footnote{Note that the fact that spec-driven development is not an entirely new concept but one that could learn from our experience in model-driven engineering has also been discussed in 
 \url{https://martinfowler.com/articles/exploring-gen-ai/sdd-3-tools.html}}). 
 Moreover, models remain a useful documentation and communication tool in any development path. This is why we say that you can have a vibe-driven approach, but, still, that approach will reamin model-based.
 
\section{Variants of vibe-driven model-based engineering}
\label{sec:variants}
As hinted above, our methodology is flexible and can be adapted to the needs of every project. This is also true within different iterations of the same project, where also other aspects as deadlines, cost, privacy,... come into play.

In fact, every iteration could navigate a different path. For instance, we could start with a more model-based approach to generate a robust first version of the system providing the core functionality in a reliable way. 
And later switch to a more vibe coding approach to, for instance, beautify the UI, as this is something predefined generators are more difficult to adapt to. Or the completely opposite scenario, where we go all in on vibe coding as a way to quickly generate prototypes useful to validate the models (and our understanding of the system requirements) with non-technical stakeholders
and then move to a model-based approach to generate trustworthy code from the validated models. 

These are just two examples of the many possible variants of this approach. 

\section{Infrastructure}
\label{sec:tool}

There is a key infrastructure element to make this proposal feasible in practice: the availability of an easy way for agents to interact with the human experts via models. In our proposal, agents are required to read, understand, create, and manipulate models \textit{from} and \textit{to} existing low-code platforms and modeling tools. 

While we can (and should) train specialized agents to become great modelers,  the agents themselves should not embed the modeling stack. 
Same as human modelers: we do not have a modeling tech stack within us; instead, we use modeling tools that expose core modeling services through different interfaces (textual, graphical, chat-based, etc.). The same applies to agents. We do not want to re-implement a full-blown modeling stack as part of each agent code. 

Instead, agents should be able to communicate with the modeling platform/s we want to use in our development project and benefit from the tool capabilities 
(e.g. to perform model validation, model rendering and many other basic model manipulation operations that are common to most modeling scenarios). 
But implementing a direct bridge between each agent and each modeling platform quickly triggers the MxN integration problem 
\footnote{The MxN integration problem refers to the challenge of connecting M different AI applications to N different external tools without a standardized approach \url{https://huggingface.co/learn/mcp-course/en/unit1/key-concepts}. 
It is a recurrent problem in information systems development that also appears in the context of vibe engineering.}.

There are now two alternative ways to bridge the gap between our modeling agents and the modeling platforms: the Model Context Protocol (MCP) and the use of agent skills.

The Model Context Protocol (MCP) \footnote{\url{https://modelcontextprotocol.io/}} is a popular open protocol that standardizes how applications provide context to agents. 
Therefore, we can add MCP support in modeling tools to bridge the gap between our modeling agents and the modeling platforms. Once a modeling platform offers an MCP server, any agent can use it to chat with it, and the platform does not need to adapt to the type of agent or the LLM used by such agent. 
Similarly, an agent embedding an MCP client can automatically discover and use any MCP server tools available in the environment without having to learn and implement code to interact with the internal modeling platform API. This allows for scenarios where agents could even use, at every step of the collaboration, a different modeling platform specialized on the type of modeling request they are working on.

As a proof of concept, we are implementing an MCP Server for the BESSER low-code platform \cite{BESSER}. 
Thanks to this MCP server, any agent can discover and use the modeling services offered by BESSER when a user (or any other agent) requests a task for which one of the BESSER services exposed in its MCP server would be a good fit. 
The actual MCP Server implementation is mostly a thin wrapper on top of the internal tool API where MCP standardizes the way the tool (in a MCP context, \textit{tool} refers to a service the agent can use as a tool to achieve something, so each modeling service would be exposed as an MCP tool)
is described (and later discovered and called) by the agent. Listing~\ref{lst:besser-mcp-example} shows a simple example of the MCP Server of BESSER exposing the creation of a new B-UML model. 
Note that the model is returned serialized. This enables the agent to keep and use the model in a future interaction if needed.
An obvious alternative would be to store the model in a database and return the ID of the model. 
Each approach has different trade-offs. 

\begin{lstlisting}[language={Python}, caption={BESSER MCP Server example}, label={lst:besser-mcp-example}]
@mcp.tool()
async def new_model(name: str) -> str:
    """Creates a new B-UML DomainModel with the specified name and returns it as base64.

    Args:
        name (str): Name of the new domain model.

    Returns:
        str: A new domain model instance as base64 string.
    """
    try:
        from besser.BUML.metamodel.structural import DomainModel  # type: ignore
    except ImportError as exc:
        raise RuntimeError(
            "BESSER library must be installed (`pip install besser`)."
        ) from exc

    # Create and return a new DomainModel instance as base64
    domain_model = DomainModel(name=name)
    return serialize_domain_model(domain_model)
\end{lstlisting}

Skills are a more recent approach to bridge the gap between agents and tools. Skills are a way to explain the set of actions that an agent can perform on a tool and how to carry them out. 
They are more like shareable workflows made of instructions and optionally code/assets, typically bundled around a SKILL.md manifest.
BESSER also comes with skills that help agents to learn and use the BESSER metamodel and the different BESSER input/output formats. Skills can also be used in complex interaction patterns 
where we want to instruct agents not to modify code generated (e.g. using a rule-based code generator) in a previous iteration of the same project (see Section \ref{sec:variants}).

The choice between MCP and skills depends on the specific needs of the project and the type of agents we are using. For instance, when you want to authenticate the agents and trace their actions, MCP is the way to go.

\section{Discussion}
\label{sec:roadmap}
So far, we have presented the key definitions and infrastructure to put in place a vibe-driven model-based approach. 
Nevertheless, there are still several challenges to address to make this approach more effective and more widely adopted. 
In what follows, we discuss some open aspects. 

\subsection{Specialized modeling agents}
While we have seen a plethora of works on inferring models from natural language descriptions (see Section \ref{sec:state-of-the-art}), most are one-shot approaches. 
The advent of agents and agentic workflows opens the door to interactive vibe modeling approaches like the one discussed here. 
But we still need to learn how to best leverage agentic capabilities and collaborate with one (or more) agent(s) to infer better models. 

Aspects like the following ones still need to be addressed:
\begin{itemize}
\item{What types of conversations and questions should the agents have with the domain experts to validate the model being inferred? Or to disambiguate and complete the natural language description?}
\item {How to train agents for the modeling domain? What type of Reinforcement Learning strategies could be useful to create specialized modeling agents?}
\item {What datasets should be created and provided to the LLMs used by the agents to improve their training (e.g. extending \cite{LopezIC22})? Or to have fine-tuned LLMs with a better understanding of the modeling concepts and tasks?}
\item {How these agents can effectively collaborate on partial models to complement and improve their own suggestions? How many agents should be involved, depending on the complexity of the model to be inferred?}
\item {How to evaluate the quality of the models inferred by the agents? And how to use that information to choose the best modeling agent for the task at hand?}
\item {How many agents should we use? And should they compete or collaborate? And if they compete, how to select the best solution from the different proposals according to a predefined governance policy\cite{AdemGovernance}}?
\end{itemize}

\subsection{Uncertainty and traceability}
Uncertainty modeling \cite{TroyaMBV21} should be considered a first-level concern. 
Indeed, when agents and LLMs are part of a model-based process, all model proposals come with a certain level of uncertainty. This confidence score should be stored together with the element. 
And for the same reason, we must be able to explain where that number came from. We need to keep full traceability of the model evolution. We should be able to explain who proposed and approved each model change.

\subsection{Adapting vibe dialogues to different user profiles}
Vibe modeling and vibe coding could be used by different types of users, from domain experts, with limited technical expertise, to software engineers with deep software engineering expertise but limited domain knowledge for the domain targeted by the system-to-be.
Each profile may prefer a different type of interaction with the agent/s. In the former, the agent should be able to explain the model in a way that is easy to understand for the domain expert. 
In the latter, the agent should focus more on bringing the domain expertise the expert engineer lacks of. This is similar to the no-code/low-code discussion, where we also have these two types of profiles, and platforms end up offering a combination of both as they are not mutually exclusive. 

\subsection{Modeling capabilities for everybody}
An implicit requirement in our approach is that the human experts have a certain level of modeling skills. At the very least, to validate and understand the models proposed by the agents.

This is not always mandatory, as in a full vibing scenario, agents could take care of creating the models that other agents will use to generate the code, while the validation of the models could be done 
by explaining such models or showing prototypes created from them to the user, i.e. an indirect validation. Still, to maximize the benefits of our approach, we would like to encourage the learning of modeling skills by everyone.

In fact, we believe this modeling (and in general abstraction \cite{bencomo2024abstractionengineering }) ability is useful beyond software development and therefore worth acquiring even for non-technical users that will only need
to occasionally participate in development projects as external collaborators.

\section{Conclusions and further work}
\label{section:conclusions-and-future-work}

This paper has introduced the concept of vibe-driven model-based engineering and how it can enable a new type of development process leveraging the best of low-code and AI to tailor the process to the needs of the 
the system to be developed and the characteristics of the user and development sceanrio. 

While this new development approach still has many shortcomings, we believe it shows promise and could contribute to reinforcing the importance of modeling in front of current trends favoring direct "vibe coding"
 of the applications with all the risks this implies for the quality of the final system. 

As further work, we plan to address the roadmap outlined above to facilitate the adoption of our approach and continue refining these ideas based on the feedback of the community and the evolution of the underlying technology.

\subsubsection*{Acknowledgements.}
This project is supported by the Luxembourg National Research Fund (FNR) PEARL program, grant agreement 16544475.

\bibliographystyle{splncs04}
\bibliography{refs}

@article{HutchinsonWR14,
  author       = {John Edward Hutchinson and
                  Jon Whittle and
                  Mark Rouncefield},
  title        = {Model-driven engineering practices in industry: Social, organizational
                  and managerial factors that lead to success or failure},
  journal      = {Sci. Comput. Program.},
  volume       = {89},
  pages        = {144--161},
  year         = {2014},
  doi          = {10.1016/j.scico.2013.03.017},
  bibsource    = {dblp computer science bibliography, https://dblp.org}
}

@article{TroyaMBV21,
  author       = {Javier Troya and
                  Nathalie Moreno and
                  Manuel F. Bertoa and
                  Antonio Vallecillo},
  title        = {Uncertainty representation in software models: a survey},
  journal      = {Softw. Syst. Model.},
  volume       = {20},
  number       = {4},
  pages        = {1183--1213},
  year         = {2021},
  doi          = {10.1007/s10270-020-00842-1},
  bibsource    = {dblp computer science bibliography, https://dblp.org}
}

@article{Booch18,
  author    = {Grady Booch},
  title     = {The History of Software Engineering},
  journal   = {{IEEE} Softw.},
  volume    = {35},
  number    = {5},
  pages     = {108--114},
  year      = {2018},
  doi       = {10.1109/MS.2018.3571234},
}

@article{BurguenoCWZ22,
  author       = {Lola Burgue{\~{n}}o and
                  Jordi Cabot and
                  Manuel Wimmer and
                  Steffen Zschaler},
  title        = {Guest editorial to the theme section on AI-enhanced model-driven engineering},
  journal      = {Softw. Syst. Model.},
  volume       = {21},
  number       = {3},
  pages        = {963--965},
  year         = {2022},
  doi          = {10.1007/s10270-022-00988-0},
  bibsource    = {dblp computer science bibliography, https://dblp.org}
}

@article{CamaraTBV23,
  author       = {Javier C{\'{a}}mara and
                  Javier Troya and
                  Lola Burgue{\~{n}}o and
                  Antonio Vallecillo},
  title        = {On the assessment of generative {AI} in modeling tasks: an experience
                  report with ChatGPT and {UML}},
  journal      = {Softw. Syst. Model.},
  volume       = {22},
  number       = {3},
  pages        = {781--793},
  year         = {2023},
  doi          = {10.1007/s10270-023-01105-5},
  bibsource    = {dblp computer science bibliography, https://dblp.org}
}

@article{LopezIC22,
  author       = {Jos{\'{e}} Antonio Hern{\'{a}}ndez L{\'{o}}pez and
                  Javier Luis C{\'{a}}novas Izquierdo and
                  Jes{\'{u}}s S{\'{a}}nchez Cuadrado},
  title        = {ModelSet: a dataset for machine learning in model-driven engineering},
  journal      = {Softw. Syst. Model.},
  volume       = {21},
  number       = {3},
  pages        = {967--986},
  year         = {2022},
  doi          = {10.1007/s10270-021-00929-3},
  bibsource    = {dblp computer science bibliography, https://dblp.org}
}

@article{RoccoRSNR25,
  author       = {Juri Di Rocco and
                  Davide Di Ruscio and
                  Claudio Di Sipio and
                  Phuong T. Nguyen and
                  Riccardo Rubei},
  title        = {On the use of large language models in model-driven engineering},
  journal      = {Softw. Syst. Model.},
  volume       = {24},
  number       = {3},
  pages        = {923--948},
  year         = {2025},
  doi          = {10.1007/S10270-025-01263-8},
  bibsource    = {dblp computer science bibliography, https://dblp.org}
}

@inproceedings{Silva24,
  author       = {Jonathan Silva and
                  Qin Ma and
                  Jordi Cabot and
                  Pierre Kelsen and
                  Henderik A. Proper},
  editor       = {Wolfgang Maass and
                  Hyoil Han and
                  Hasan Yasar and
                  Nicholas J. Multari},
  title        = {Application of the Tree-of-Thoughts Framework to LLM-Enabled Domain
                  Modeling},
  booktitle    = {Conceptual Modeling - 43rd International Conference, {ER} 2024, Pittsburgh,
                  PA, USA, October 28-31, 2024, Proceedings},
  series       = {Lecture Notes in Computer Science},
  volume       = {15238},
  pages        = {94--111},
  publisher    = {Springer},
  year         = {2024},
  doi          = {10.1007/978-3-031-75872-0\_6},
  bibsource    = {dblp computer science bibliography, https://dblp.org}
}

@inproceedings{chen2023automateddomainmodeling,
	address = {Västerås, Sweden},
	title = {Automated {Domain} {Modeling} with {Large} {Language} {Models}: {A} {Comparative} {Study}},
	isbn = {9798350324808},
	shorttitle = {Automated {Domain} {Modeling} with {Large} {Language} {Models}},
	doi = {10.1109/MODELS58315.2023.00037},
	booktitle = {2023 {ACM}/{IEEE} 26th {International} {Conference} on {Model} {Driven} {Engineering} {Languages} and {Systems} ({MODELS})},
	publisher = {IEEE},
	author = {Chen, Kua and Yang, Yujing and Chen, Boqi and Hernández López, José Antonio and Mussbacher, Gunter and Varró, Dániel},
	month = oct,
	year = {2023},
	pages = {162--172},
}

@article{Lola24,
  author       = {Lola Burgue{\~{n}}o and
                  Davide Di Ruscio and
                  Houari A. Sahraoui and
                  Manuel Wimmer},
  title        = {The Past, Present, and Future of Automation in Model-Driven Engineering},
  journal      = {CoRR},
  volume       = {abs/2405.18539},
  year         = {2024},
  doi          = {10.48550/ARXIV.2405.18539},
  eprinttype    = {arXiv},
  eprint       = {2405.18539},
  bibsource    = {dblp computer science bibliography, https://dblp.org}
}

@article{fill_conceptual_2023,
	title = {Conceptual {Modeling} and {Large} {Language} {Models}: {Impressions} {From} {First} {Experiments} {With} {ChatGPT}},
	copyright = {Creative Commons Attribution Share Alike 4.0 International},
	shorttitle = {Conceptual {Modeling} and {Large} {Language} {Models}},
	doi = {10.18417/EMISA.18.3},
	language = {en},
	urldate = {2023-10-19},
	journal = {Enterprise Modelling and Information Systems Architectures (EMISAJ)},
	author = {Fill, Hans-Georg and Fettke, Peter and Köpke, Julius},
	month = apr,
	year = {2023},
	note = {Artwork Size: 3:1-15 Pages
Publisher: Enterprise Modelling and Information Systems Architectures (EMISAJ)},
	pages = {3:1--15 Pages},
}

@inproceedings{chaaben_towards_2023,
	address = {Melbourne, Australia},
	title = {Towards using {Few}-{Shot} {Prompt} {Learning} for {Automating} {Model} {Completion}},
	isbn = {9798350300390},
	doi = {10.1109/ICSE-NIER58687.2023.00008},
	language = {en},
	urldate = {2023-09-15},
	booktitle = {2023 {IEEE}/{ACM} 45th {International} {Conference} on {Software} {Engineering}: {New} {Ideas} and {Emerging} {Results} ({ICSE}-{NIER})},
	publisher = {IEEE},
	author = {Chaaben, Meriem Ben and Burgueño, Lola and Sahraoui, Houari},
	month = may,
	year = {2023},
	pages = {7--12},
}

@inproceedings{lowmodeling,
  author       = {Jordi Cabot},
  editor       = {Hans{-}Georg Fill and
                  Francisco Jos{\'{e}} Dom{\'{\i}}nguez Mayo and
                  Marten van Sinderen and
                  Leszek A. Maciaszek},
  title        = {Low-Modeling of Software Systems},
  booktitle    = {Software Technologies - 18th International Conference, {ICSOFT} 2023,
                  Rome, Italy, July 10-12, 2023, Revised Selected Papers},
  series       = {Communications in},
  volume       = {2104},
  pages        = {19--28},
  publisher    = {Springer},
  year         = {2023},
  doi          = {10.1007/978-3-031-61753-9\_2},
  bibsource    = {dblp computer science bibliography, https://dblp.org}
}

@article{TerragniVRB25,
  author       = {Valerio Terragni and
                  Annie Vella and
                  Partha Roop and
                  Kelly Blincoe},
  title        = {The Future of {AI}-Driven Software Engineering},
  journal      = {ACM Trans. Softw. Eng. Methodol.},
  volume       = {34},
  number       = {5},
  articleno    = {120},
  year         = {2025},
  month        = may,
  publisher    = {Association for Computing Machinery},
  address      = {New York, NY, USA},
  issn         = {1049-331X},
  doi          = {10.1145/3715003}
}

@inproceedings{BESSER,
  author       = {Iv{\'{a}}n Alfonso and
                  Aaron David Conrardy and
                  Armen Sulejmani and
                  Atefeh Nirumand and
                  Fitash Ul Haq and
                  Marcos Gomez{-}Vazquez and
                  Jean{-}S{\'{e}}bastien Sottet and
                  Jordi Cabot},
  editor       = {Han van der Aa and
                  Dominik Bork and
                  Rainer Schmidt and
                  Arnon Sturm},
  title        = {Building {BESSER:} An Open-Source Low-Code Platform},
  booktitle    = {Enterprise, Business-Process and Information Systems Modeling - 25th
                  International Conference, {BPMDS} 2024, and 29th International Conference,
                  {EMMSAD} 2024, Limassol, Cyprus, June 3-4, 2024, Proceedings},
  series       = {Lecture Notes in Business Information Processing},
  volume       = {511},
  pages        = {203--212},
  publisher    = {Springer},
  year         = {2024},
  doi          = {10.1007/978-3-031-61007-3\_16},
  bibsource    = {dblp computer science bibliography, https://dblp.org}
}

@article{RuscioKLPTW22,
  author       = {Davide Di Ruscio and
                  Dimitrios S. Kolovos and
                  Juan de Lara and
                  Alfonso Pierantonio and
                  Massimo Tisi and
                  Manuel Wimmer},
  title        = {Low-code development and model-driven engineering: Two sides of the
                  same coin?},
  journal      = {Softw. Syst. Model.},
  volume       = {21},
  number       = {2},
  pages        = {437--446},
  year         = {2022},
  doi          = {10.1007/S10270-021-00970-2},
  bibsource    = {dblp computer science bibliography, https://dblp.org}
}

@inproceedings{vibemodeling,
  author       = {Jordi Cabot},
  title        = {Vibe Modeling: Challenges and Opportunities},
  booktitle    = {Advances in Conceptual Modeling - {ER} 2025 Workshops, FCM, CMLS,
                  LLM4Modeling, OntoCom, and QUAMES, Poitiers, France, October 20-23,
                  2025, Proceedings},
  series       = {Lecture Notes in Computer Science},
  pages        = {105--118},
  publisher    = {Springer},
  year         = {2025},
  url          = {https://doi.org/10.1007/978-3-032-08620-4\_7},
  doi          = {10.1007/978-3-032-08620-4\_7},
  bibsource    = {dblp computer science bibliography, https://dblp.org}
}

@book{cabot2024lowcode,
  title={The low-code handbook: Learn how to unlock faster and better software development with low-code solutions},
  author={Cabot, Jordi},
  year={2024},
  month={October},
  publisher={Self-published},
  isbn={978-9998778511},
  asin={B0DK2ZPTQX},
  pages={145},
  edition={1st},
  language={English},
  url={https://lowcode-book.com/}
}

@article{manifestmodeling,
author = {Lukyanenko, Roman and Samuel, Binny and Tegarden, David and Larsen, Kai and Jabbari, Araz and Abd El Aziz, Rasha and Almeida, Joao and Amrollahi, Alireza and Beard, Jon and Bellatreche, Ladjel and Bork, Dominik and Brocke, Jan vom and Cabot, Jordi and Castellanos, Arturo and Gerber, Aurona and Green, Peter and Grover, Andrea and Guizzardi, Giancarlo and Hertelendy, Attila and Zhao, J.},
year = {2025},
month = {12},
pages = {},
title = {Manifesto for Software Development and Modeling in the Age of Artificial intelligence},
journal = {SSRN Electronic Journal},
doi = {10.2139/ssrn.5881104}
}

@inproceedings{SilvaMCKP25,
  author       = {Jonathan Silva Mercado and
                  Qin Ma and
                  Jordi Cabot and
                  Pierre Kelsen and
                  Henderik A. Proper},
  editor       = {Dominik Bork and
                  Roman Lukyanenko and
                  Shazia Sadiq and
                  Ladjel Bellatreche and
                  Oscar Pastor},
  title        = {Towards Human-in-the-Loop LLM-Enabled Domain Modeling},
  booktitle    = {Conceptual Modeling - 44th International Conference, {ER} 2025, Poitiers,
                  France, October 20-23, 2025, Proceedings},
  series       = {Lecture Notes in Computer Science},
  pages        = {127--145},
  publisher    = {Springer},
  year         = {2025},
  url          = {https://doi.org/10.1007/978-3-032-08623-5\_7},
  doi          = {10.1007/978-3-032-08623-5\_7},
  bibsource    = {dblp computer science bibliography, https://dblp.org}
}

@inproceedings{AdemGovernance,
  author       = {Adem Ait and
                  Gwendal Jouneaux and
                  Javier Luis C{\'{a}}novas Izquierdo and
                  Jordi Cabot},
  title        = {Towards Automated Governance: {A} {DSL} for Human-Agent Collaboration
                  in Software Projects},
  booktitle    = {40th {IEEE/ACM} International Conference on Automated Software Engineering,
                  {ASE} 2025, Seoul, Korea, Republic of, November 16-20, 2025},
  pages        = {3891--3895},
  publisher    = {{IEEE}},
  year         = {2025},
  url          = {https://doi.org/10.1109/ASE63991.2025.00337},
  doi          = {10.1109/ASE63991.2025.00337},
  bibsource    = {dblp computer science bibliography, https://dblp.org}
}

@misc{bencomo2024abstractionengineering,
      title={Abstraction Engineering}, 
      author={Nelly Bencomo and Jordi Cabot and Marsha Chechik and Betty H. C. Cheng and Benoit Combemale and Andrzej Wąsowski and Steffen Zschaler},
      year={2024},
      eprint={2408.14074},
      archivePrefix={arXiv},
      primaryClass={cs.SE},
      url={https://arxiv.org/abs/2408.14074}, 
}

\end{document}